# Ontology Enrichment by Extracting Hidden Assertional Knowledge from Text

Meisam Booshehri+,1, Abbas Malekpour+, Peter Luksch+

+Department of Distributed High Performance Computing, Institute of Computer Science, University of Rostock,
Rostock, Germany

Kamran Zamanifar++, Shahdad Shariatmadari+++
++Faculty of Computer Engineering,
Najfabad Branch, Islamic Azad University , Najafabad, Iran
+++Faculty of Computer Engineering,
Shiraz Branch, Islamic Azad University , Shiraz, Iran

*Abstract*—In this position paper we present a new approach for discovering some special classes of assertional knowledge in the text by using large RDF repositories, resulting in the extraction of new non-taxonomic ontological relations. Also we use inductive reasoning beside our approach to make it outperform. Then, we prepare a case study by applying our approach on sample data and illustrate the soundness of our proposed approach. Moreover in our point of view current LOD cloud is not a suitable base for our proposal in all informational domains. Therefore we figure out some directions based on prior works to enrich datasets of Linked Data by using web mining. The result of such enrichment can be reused for further relation extraction and ontology enrichment from unstructured free text documents.

## I. INTRODUCTION

Information Extraction is categorized into three tasks [21]: Named Entity (in a nutshell NE) Recognition, Named Entity Disambiguation and Relation Extraction. Actually recognition of named entities deals with finding textual mentions of entities which belong to a set of categories including persons, organizations, places, etc. In disambiguation of named entities we relate the mentions of entities in the text to an external entity. Finally in relation extraction process we extract semantic relations between predefined named entities.

By applying relation extraction process we can convert unstructured data (we mean free texts) into structured data. This makes it possible to apply so many algorithms in the field of data mining, question answering and semantic web [21]. To the best of our knowledge current methods for relation extraction are classified as follows: Manual relation extraction methods, supervised methods, semi-supervised methods and unsupervised methods.

With emerging the web of Linked Data, so many researchers have tried to make use of its potential benefits [1, 2, 16, 17 and 30]. Also we believe that Linked Data has hidden potential benefits. There are some approaches which uses Linked Data to discover the relations between NE pairs in a text [3].

On the other hand there are different systems that automatically generate ontology from text. There are many researchers who are working on Ontology Learning layers. To date, researches have resulted in creation of an 8-layer Ontology Learning Stack. The layers of this Stack are: terms layer, Synonyms layer, Concept Formation layer, Concept Hierarchy Layer, Relations Layer, Axiom Schemata Layer and General Axioms Layer [13].

In this paper we introduce an approach which could be done after the ontology learning tasks are done. In this approach we try to find hidden relations in input texts by using Linked Data. In other words we try to discover a special class of assertional knowledge, resulting in the extraction of new non-taxonomic ontological relations. Some components of such knowledge are invisible in the text so we use Linked Data to make it appear. Although this approach has the power to enrich instances related to the concepts of the ontology. Actually we see Linked Data as a huge giant global database that can be used to enrich the ontology extracted from a text both in Schema layer and instance layer.

There are some similarities and differences between our proposed approach for using Linked Data to enrich an ontology and relation extraction methods which uses Linked Data to annotate resources in a text. So we present a comparative study and mention some critiques on existing relation extraction methods in the following sections.

The remaining sections are organized as follows. The second section deals with background and related work. The third section describes invisible meaning and defines a new problem. The fourth section describes a new approach for enriching an ontology. The fifth section presents a comparative study on co-occurrence limitations of NE pairs in different methods. The sixth section comes up with discussions. Finally the seventh section is the conclusion and eighth section is future work.

## II. RELATED WORK

### A. Relation Extraction Methods

In [23] and [26] two of the earlier approaches for relation extraction from biological text documents have been proposed. In these approaches some relations are extracted based on a set of rules which have been created manually. In supervised relation extraction methods some predefined relations are considered among named entities. Learning based on SVM and kernel functions are examples of such approach [27 and 28]. Also in [21] a multi instance learning method has been proposed that is considered to be a supervised method. Unsupervised methods usually work based on clustering techniques. In [29] an unsupervised method has been proposed which is based on clustering for discovering the relations among NE pairs. In [22] a fully unsupervised method for web mining has been proposed with which we can extract the relations that one of their arguments is a predefined concept.

---

1 Corresponding Author at : Department of Distributed High Performance Computing, Institute of Computer Science, University of Rostock, Rostock, Germany; Email: m_booshehri@sco.iaun.ac.ir

### B. Automatic Ontology Creation From Text

Different systems for automatic ontology creation have been constructed up to now which cover different layers of the ontology learning stack [13, 18, 19 and 20]. We just mention some few systems here. Text2Onto covers the first five layers of ontology learning stack [6 and 7]. AOEN covers only the axiom schemata layer. HASTI [4] covers the terms layer, concept hierarchy layer, relations layer, general axioms layer. OntoLearn covers the first five layers. ATRACT covers the first three layers. Paramenidens covers the first two layers [10 and 13] and etc. To the best of our knowledge, no system has ever been constructed to cover all the eight layers of the ontology learning stack. And no system has ever used Linked Data to improve the process of ontology learning from text. Also there has been no effort to extract the Implied Information (hidden assertional knowledge) from texts which results in new ontological relations as we will talk about it in fourth section.

### C. Resource annotation and Relation Extraction by Using Linked Data

[5] presents and evaluate two existing word sense disambiguation approaches which are adopted to annotate text with several popular Linked Open Data datasets. [3] utilizes Linked Data to generate semantic annotations for frequent patterns extracted from textual documents.

### III. INVISIBLE MEANING AND DEFENITION OF A PROBLEM

Here we introduce some special classes of knowledge which can be useful in ontology learning or relation extraction process. We believe that such classes of knowledge could be discovered only by data mining methods because there is weak information about such knowledge in the text and we can reach the lost rings of it by data mining process both in the traditional web and Linked Data. The original concept of such class of knowledge derives from "discourse analysis" and "pragmatics" in linguistics. An important characteristic these two practices share is, according to Yule, the study of "invisible meaning": "how we recognize what is meant even when it isn't actually said or written" [11]. Yule mentions a number of devices we use to discover these invisible meanings, amongst them "context" and "inference." To draw an analogy, a context would be the information domain we are dealing with, which makes clear where in its possibly wide range of meaning a word is functioning .Actually we can use this concept for word sense disambiguation. An inference, though, would be any ontological relation which is implicit in the text (from which the ontology is created) because only some components of it appear. Based on this discussion we define three classes of knowledge. We consider the knowledge containing a relation between two named entities equal to an RDF triple which consists of a subject, a predicate and an object.

Definition. 1. **One-component-in-text Knowledge:** It is the knowledge which just one component (subject or object) of it has appeared in the text. Suppose that the concept "country" has appeared in the text. Now every knowledge in real world that this concept can take part in, is some one-component-in-text knowledge in viewpoint of the user that reads the text. Or suppose the word "France" which is an instance of the concept "country", has appeared in the text. The complete set of relations in the real world, in which the word "France" is present, is the same set of one-component-in-text Knowledge starting from the word "France". A person who reads a text has to be familiar with some one-component-in-text knowledge about a specific word appeared in the text, that is a user that see a word in a text should know some possible meanings of that word. Such knowledge about words in a text helps the user to understand the text.

Definition 2. **Two-component-in-text Knowledge:** It is the knowledge that exactly two components (subject and object) of it have appeared in the text. The components may be positioned far from each other in the text. In this case no predicate has been mentioned for the knowledge in the text. We explain it with a scenario. Suppose the person A is a professor of computer science in the university X and the person B has finished his Ph.D. level in university X under the supervision of person A. On the other hand we have a text about ISWC conference from which we want to extract some relations. In this text the names of general chair, track chairs and some other people have been mentioned. Now suppose that the person A is the general chair of the conference and the person B is one of the track chairs of the conference and there is no knowledge in the text insisting that the person A has been the supervisor of the person B. With these assumptions, learning such knowledge that "the Person A has been the supervisor of person B" from this text is possible with current relation extraction methods only in the case of using data mining methods which use a background knowledge such as web content to extract such relations. Such assertional knowledge is called two-component-in-text knowledge.

Definition 3. **Three-component-in-text knowledge:** It is the knowledge which all three parts of it have appeared in the text. It is clear that the subject and the object of this knowledge could have other predicates not mentioned in the text. For more, remember the scenario we mentioned for explaining two-component-in-text knowledge except that there is at least one sentence in the text which contains all three parts of the knowledge. Such knowledge could be extracted from text by using current methods of ontology learning from text without need to any background knowledge about the knowledge components.

**Problem Definition:** Given a text we want to know how we can make use of Two-component-in-text Knowledge and Three-component-in-text Knowledge to enrich the ontology created from that text. We propose a method which can uses such knowledge to enrich the ontology created from text by using Linked Data.

## IV. ENRICHING THE INTERMEDIATE ONTOLOGY BY USING LINKED DATA

In this section the proposed approach is described. Actually it is a step that can be done after ontology learning tasks. The task of this approach is to enrich the output ontology extracted from every combination of previous 8 layers. To realize such a task, we present a new algorithm which uses Linked Data to enrich the ontology created from text. After that we show the soundness of our algorithm by bringing real examples which use current real Linked Data. We have prepared high level descriptions of our algorithm as follows in the current section.

The idea is that the learning process begins with respect to the ontology learning stack. Indeed by processing input text, an intermediate ontology is created. This intermediate ontology is equivalent to the output ontology of tools such as Text2Onto [6] which use almost the best techniques in the field of ontology learning. Now we can send this intermediate ontology to the new approach to be enriched by using Linked Data database.

The proposed approach enriches the non-taxonomic relations by processing the corresponding instances of the ontology concepts. A high level description of the methodology that we propose to enrich the intermediate ontology in the new approach is as follows.

1- Intermediate Ontology Extraction by using techniques in previous 8-layers of the ontology learning stack.

2- Forming the set of instances of intermediate ontology and computing the Cartesian of this set. These instances are components of some *two-component-in-text knowledge* or some *three-component-in-text knowledge* existing in the text. Here we can omit some ordered pairs in the Cartesian set. For example we may omit the ordered pairs with equal elements. Also we may omit every ordered pair which its elements are positioned far from each other in the text. It is based on the idea that if two instances are positioned far from each other in the text it means that there is a weak relation between them [5]. In fifth section we have prepared a comparative study on this subject.

3- Now we pass the Cartesian set to our algorithm to find the new suitable predicates related to the domain of the text for every member of the set.

4- After finding the suitable predicates, the algorithm relates the instances to the corresponding concepts in the schema layer of the intermediate ontology.

5- In this step we should review the ontology and check some relations such as transitivity relations to optimize the schema layer of the ontology. Also we can use inductive reasoning to help enriching process.

The proposed steps are as follows.

```
Input:
A={The Cartesian set of instances existing
in the instance layer of intermediate
ontology}
= {OP₁ , OP₂, …., OPₙ*ₙ} =
{(subject₁,object₁),…, (subjectₙ*ₙ, objectₙ*ₙ)}
CorrespondingConceptₙ }
LD: Linked Data database
Maxtime: maximum time preferred to search
for RDF pages in Linked Data Database
Output:
An Enriched Ontology Named O
Pseudo-Code:
  1. for(int k=0;k<n*n; k++)
  2. {
  3. att=FindPredicate (LD, A[i]["subject"],
     A[i]["Object"])
  4. if (att != NULL)
  5. add the
     Assertional_knowledge"(A[i]["subject"],
     att , A[i]["Object"])" to Ontology O

  6. add the rule"(corresponding Concept
     Of(A[i]["subject"]), att , corresponding
     Concept Of (A[i]["Object"]))" to
     Ontology O

  7. }
```

As you see there are two functions used in this algorithm. We explain the algorithm as comes below:

**FindPredicate function**: this function has a formal parameter named "Alpha". This parameter holds the similarity value that user considers as an acceptable factor. The Pseudo-Code of this function has come below.

```
  1. FindPredicate (LD, e1, e2, Alpha)
  2. {
  3. RDFPages=
     searchRDFWithSimilarityCheck(LD,e1,Maxti
     me)
  4. for each(RDFtriple in RDFPages)
  5. {
  6. if(RDFtriple.Object=e2)
  7. if(ContextSimilarity(RDFtriple.Object,
     e2)> Alpha)
  8. return RDFtriple.Predicate
  9. }
  10. }
```

Note that *searchRDFWithSimilarityCheck* function searches for all RDF triples which their subjects' name are equal to *e1*'s name with considering the variable *Maxtime* which is the threshold of search time. After finding such triples, some are chosen with respect to the Similarity of e1 and subjects of found RDF triples in Linked Data. Actually *e1* is the first instance which is our current subject to search for, and e2 is the second instance which is our current object. We check the similarities by using ContextSimilarity Function.

**ContextSimilarity Function**: The Pseudo-Code of this function is as comes below. We mention and use exactly the same algorithm with the same notation mentioned in [5]. Also there are discussions about similarity reckoning in [15] and [16]; however we won't get involved in this subject in the current paper and we just accept one of the existing methods to compute similarity as follows. Also we must take care about the performance of the method.

```
ContextSimilarity (resource, w_a) returns Similarity
   1. Similarity=0
   2. NR= GetNeighborhoodResources(resource)
   3. CW= GetContext(w_a)
   4. for i=1 to size(NR) do
   5. CS= sim_cos(NR[i], CW)
   6. Similarity= Similarity+CS
   7. end for
   8. return Similarity
```

In general the objective of our algorithm is enriching non-taxonomic relations by standing on the shoulder of instance layer formed in the intermediate ontology. The algorithm searches for relations (= predicates) between instances of the ontology layer in the Linked Data. After finding suitable predicates, these predicates are related to the corresponding concepts in the intermediate ontology. The reason for using the term "suitable predicate" is that we are not going to add semantic relations between our recognized instances in another domains or datasets which are not related to our ontology domain. Capability of adding such relations don't result in quality improvement of ontology. Actually our objective is not creating an ontology that covers every relation in every domain. One of the conditions we seek is domain matching, that is, we add the found predicate in Linked Data to our intermediate ontology in the case that the domain of our text is the same as the domain of the "subject" and "object" of the current RDF triple in Linked Data. Recognizing this identity is related to the Dataset that we choose in Linked data. One of the algorithms that is used for recognizing the identity of the domain of a resource in the text and the domain of the similar resource in the Linked Data is Context Similarity. Many of LOD datasets such as Freebase, DBpedia, Wordnet and OpenCyc connect a comment to their resources. For example in DBpedia, comments about every resource are found under rdfs:comment. In context similarity algorithm similarity of "the comments of a resource in Linked Data" and "related concepts of a resource in the text" is determined by using statistical techniques. So we use this algorithm as a function in our algorithm.

To illustrate the soundness of our algorithm we put forward an example in the geographical domain. Consider the following text:

"Geography is the science that deals with the study of the Earth. In Geography we discuss geographical entities such as Natural Geographical Entities and Inhabited Geographical Entities. Generally in geography we talk about cities, countries and other inhabited geographical entities. A country is a geographical region that contains smaller regions called "city". In political point of view, one of the large cities which are located in a country is chosen to be the capital of the country. Therefore, every country has a capital city. Here we introduce some Geographical Entities briefly.

Germany is a country in Western and Central Europe. The Capital and largest city of Germany is Berlin. One of the famous cities which are located in Germany is Stuttgart.

Another example is Iran, officially the Islamic Republic of Iran, which is a country in Central Eurasia and Western Asia. It is a country of particular geostrategic significance due to its location in the Middle East and central Eurasia.

Other geographical entities that we discuss in geography are Natural Geographical Entities such as mountains, rivers, forests. For example The Zugspitze, with a peak of 2,962 meters above sea level, is the highest mountain in Germany. There is also a forest named Black Forest located in Germany. There are well-known rivers such as Neckar which flow through Germany, passing different cities such as Stuttgart. Neckar is 367 km long. Zard kuh, as another example, is a mountain in Iran.

The Shatt al-Arab is a river in Southwest Asia. At first the Tigris and the Euphrates join in Iraq and the Karun river joins the waterway from Iranian side and as a result The Shatt al-Arab is formed."

Now if we analyze this text according to current methods and semantic patterns such as Hearst pattern, an ontology is created as shown in Figure 1. This ontology has been created based on existing three-component-in-text knowledge in the text.

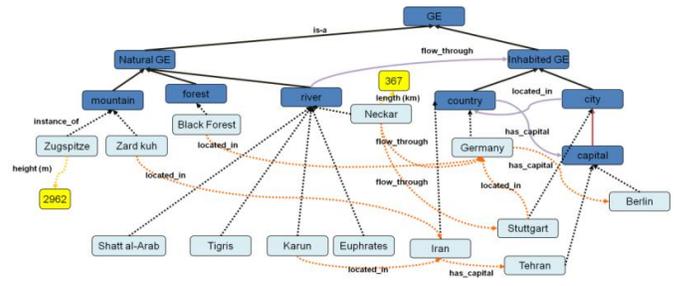

Figure 1

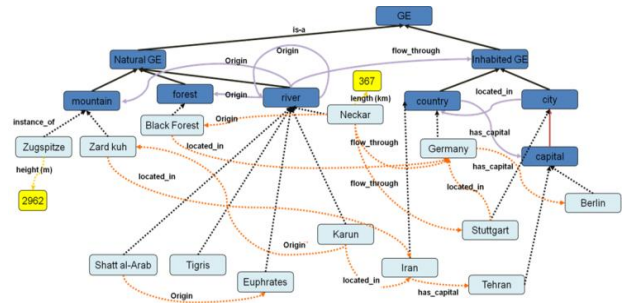

Figure 2

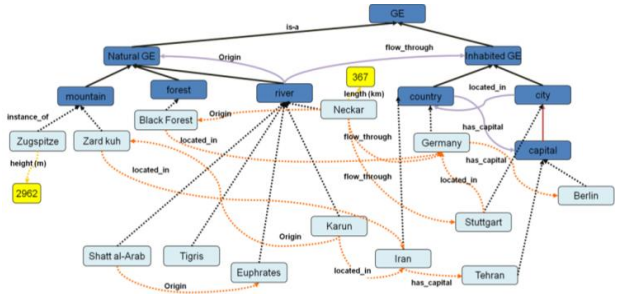

Figure 3

We consider this ontology as an intermediate ontology which is the base of our examples in the following sections.

### A. Enriching the intermediate ontology : A case study

In this section we show the enriching process of the intermediate ontology created from text through an example. In [13] binary relations are introduced and a notation is chosen to describe the relations. We use the same notation in the whole paper. Suppose a relation **r**. Every relation has a domain shown with **dom(r)** and range shown with **range(r)**. For example suppose a geographical ontology that has concepts such as river, city, country, Geographical entity (in a nutshell GE) etc. A relation such as: "pass_through (dom: river, range: GE)" means that: "An entity of the type river can pass_through an entity of the type GE".Now consider the ontology shown in Fig. 1. We name the set of instances in the ontology as B.

$$B = \{Zugspitze, Zard\ kuh, Black\ forest,\ Neckar,\\
Euphrates, Tigris, Karun, Berlin, Stuttgart\\
Tehran, Germany, Iran, Shatt\ al-Arab\}$$

Now we should compute the Cartesian of set B as follows:

$$A = B \bowtie B\\
= \{(Zugspitze, Zugspitze), (Zugspitze, Zard\ kuh), ...\}$$

Also in our intermediate ontology we have the following set:

$$NGE = Natural\ Geographical\ Entity\\
= \{mountain, forest, river\}$$

Generally in this example the number of members of set A is 13*13=169. We discuss three ordered pair of the set A which we have found suitable predicates for them. To find suitable predicates we have used <u>FactForge.net</u> . We have shown the ordered pairs and the corresponding RDF triple that we have found for each of them as follows.

$$(Neckar, Black\ Forest)\\
\rightarrow (Neckar, Origin, Black\ Forest)$$

$$(Shatt\ al-Arab, Euphrates)\\
\rightarrow (Shatt\ al-Arab, Origin, Euphrates)$$
$$(Karun, Zard\ Kuh) \rightarrow (Karun, Origin, Zard\ Kuh)$$

By processing RDF triples which we have found, we can conclude the following rules to add to the intermediate ontology. As a result our ontology would be as is shown in Figure 2.

$$Origin(dom: river, \mathbf{range:}\ forest)(1)$$
$$Origin(dom: river, rnage: river)(2)$$
$$Origin(dom: river, range: mountain)(3)$$

Since in the above ontology we have the following axiom:

$$\forall x \in NGE \rightarrow Origin(dom: river, range: x)(4)$$

So we can conclude that the following equation holds:

$$Origin(dom: river, range: river)$$
$$Origin(dom: river, range: mountain)$$
$$Origin(dom: river, range: forest)$$
$$\leftrightarrow$$
$$Origin(dom: river, range: Natural\ GE)(5)$$

Therefore the ontology changes as is shown in Figure 3.

### B. Inductive reasoning to help enriching process

Reasoning is the process of arriving at conclusions from evidence. Inductive Reasoning is reasoning from particular facts [leading] to general principles. In Inductive Reasoning, we don't assert that something is true; it is probably more true than not. The larger the number of specific instances, the more certain is the generalization. Actually inductive reasoning is the reasoning from specific cases to more general, but uncertain, conclusions. Another type of reasoning is deductive reasoning which is reasoning from general premises, which are known or presumed to be known, to more specific, certain conclusions. Generally a mathematical theorem is created as follows. At first we should observe around the world or actually among the members of a set in real world to find a hidden relation. The whole set of such relations indicate that a hypothesis may be true based on inductive reasoning. Thus by using deductive reasoning we can prove this hypothesis.

In accordance with the following scenario inductive reasoning could prepare a ground to find new ontological knowledge to add to intermediate ontology.

Remember the case study mentioned in previous section. Suppose that by searching in Linked Data in the first step in a limited time we reach the relations 1 and 2. And we don't reach a relation such as relation No. 3. Now assume that (1) and (2) holds. Based on inductive reasoning we can result that in the set NGE, the relation **(4)** may hold. (1) And (2) are evidences of this claim.

$$Origin(dom: river, range: forest)$$
$$Origin(dom: river, rnage: river)$$
$$NGE = \{forest, river, mountain\}$$

$$\xrightarrow{inductive\ reasoning}$$

$$\forall x \in NGE \quad Origin(dom: river, range: x) \quad \textbf{(4)}$$

To prove this we can search Linked Data again just by using a simple sparql query. By proving this claim it is clear that the intermediate ontology would be more enriched.

$$Origin(dom: river, range: mountain) \quad (€)$$

Suppose that by searching Linked Data we can find such assertional knowledge as follows:

$$origin (dom: karun, range: zardkuh)$$

This knowledge insists that the relation (€) holds. Therefore we can say that the hypothesis has been proved in the current space of our ontology.

## V. A Comparative study on Co-occurrence LiMitations Of The Named Entities

Many of relation extraction methods limits the co-occurrence of the words within a sentence and the NE pairs that are seen to occur in a sentence is assumed to be co-occurred; however there is no limit for co-occurrence of words in real world. But the bigger space we consider for co-occurrence of two words, the more time we need to search for the relations of words because of increase in number of NE pairs. Many of such relations may not be useful in our application. But we believe that considering the co-occurrence of two words as occurring in a sentence may result in the obsolescent of some useful information amongst that two-component-in-text knowledge as we described in the related scenario.

In our method for enriching the intermediate ontology we extract hidden assertional knowledge from text by using Linked Data. In the case of our algorithm, hidden knowledge is discovered while two following conditions are established:

- "Subject" and "Object" of an RDF triple (= our target knowledge) exist in the text.
- Our target Linked Data has at least one RDF triple with the same "subject" and "object" ,in the same domain.

We think that Linked Data consist of assertional knowledge (also called facts). Therefore our proposed approach in this paper is an *approach for extracting some hidden assertional knowledge from text by using proper Linked Data dataset which results in achieving new Ontological Knowledge*.

As cleared above in our method we don't pay attention to the co-occurrences of the words in the text; we just compute the Cartesian set as we described in previous section and search for the suitable predicate for the members of the Cartesian set. This is because we think that classes of an ontology may have strong association relationships, thus resulting in strong relations between instances of the ontology classes. As you see in the case study the words "Zard Kuh" and "Karun" are not co-occurred in a sentence in the text; however combination of these two words give us proper assertional knowledge resulting in proper ontological knowledge.

Totally we think that from the word co-occurrence aspect our method for relation extraction results in lower obsolescent of information in comparison to existing relation extraction methods which we introduced in related work section. Evaluation of this claim would be one our future works.

## VI. Obsolescent Of Information In Linked Data and Enriching Datasets Of Linked Data

Linked Data does not have rich contents in all informational domains. Recently, some statistics have been presented that show the growth of Linked Data from June 2009 to Nov. 2010. The growth has been 300%. True that such percent may sound so huge, but the amount of structured data existing in Linked Data in comparison to the amount of unstructured data existing in traditional web or in comparison to the number of relations between the words in real world is very small. Actually almost 90 percent of data in human being world are created and maintained in an unstructured form. For example web pages, emails, technical documents, corporate documents, books, etc. are kept in an unstructured form. This study shows the obsolescent of information in Linked Data. So some suitable frameworks must be provided to accelerate the growth rate of information in Linked Data more and more.

In [22] a fully unsupervised approach for relation extraction by web mining has been proposed with which we can extract the relations that one of their arguments is a predefined concept. Actually we think that it can be used in order to discover a set of *one-component-in-text knowledge* according to the existing text. Also in our point of view such methods can make use of *one-component-in-text knowledge* for automating the process of enriching the datasets of Linked Data by web mining.

## VII. Discussion

Generally, the philosophy of our proposed approach to enrich the intermediate ontology created from text is based on two grounds. The first ground is the notion of Linked Data and LOD formation to realize semantic web. Generally, since Liked Data "makes the web appear as one giant huge global database," we could use this database to find new predicates related to the concepts in the intermediate ontology. The quotation has not been completely realized yet.

Our second ground derives from "discourse analysis" and "pragmatics" in linguistics. An important characteristic these two practices share is, according to Yule, the study of "invisible meaning": "how we recognize what is meant even when it isn't actually said or written" [11]. Yule mentions a number of devices we use to discover these invisible meanings, amongst them "context" and "inference." To draw an analogy, a context would be the information domain we are dealing with, which makes clear where in its possibly wide range of meaning a word is functioning. An inference, though, would be any ontological relation which is implicit in the text (from which the ontology is created) because only some components of it appear.

We believe that Linked Data has potential benefits. A tangible example is using Linked data in ontology learning processes. Although datasets of Linked Data such as DBpedia are believed to be a set of best practice for exposing, sharing, and connecting pieces of *data*, *information*, and *knowledge* on the Semantic Web using URIs and RDF [1,2, 16, 17 and 30], we use another definition for describing Linked Data. In our point of view, Linked data is a type of collective knowledge which must be the result of collective wisdom and experience. This collective knowledge which has appeared in LOD cloud in evolution. So it becomes clear that every method in ontology engineering which is related to Linked Data would inherit dynamism from the nature of Linked Data. In other words, Linked data dynamism propagates itself inside the methods which use Linked Data as a reference database.

In any text, there is some hidden information as against evident information. Evident information is all that the author has himself expressed quite explicitly and consciously. Hidden information, on the other hand, is all that is only implied in a text. The process by which such hidden or implied information (hidden assertional knowledge) is made apparent is "deductive inference" [12]. We argue that using Linked Data in ontology learning processes can make use of inferences to reveal such hidden information and to infer from them specific ontological relations which would not be otherwise extracted. To better illustrate this point, we draw your attention to the following example:

"At first the Tigris and the Euphrates join in Iraq and the Karun river joins the waterway from Iranian side and as a result The Shatt al-Arab is formed. The Shatt al-Arab is a river in Southwest Asia of some 200 km (120 mi) length."

In the above passage it is clear that three rivers join to form the Shatt al-Arab. But the piece of information, and accordingly the ontological relation, which is not explicit is that "a river can originate from another river." We consider it as a piece of hidden information. With an implied piece of information some components of the ontological relation we wish to infer do appear in the text. For example, the "subject" and the "object" of an RDF triple are analogous to the components just mentioned. Using our method results in the revealing of such hidden information. For instance, in the example mentioned in the fourth section, the following relations have been discovered:

$$origin(dom: river, range: forest)(1)$$
$$origin(dom: river, range: river)(2)$$
$$origin(dom: river, range: mountain)(3)$$

The ontology can be even further optimized as the following relation has been resulted from three discovered relations mentioned above:

$$origin(dom: river, range: Natural GE)$$

To define hidden information more clearly, we make use of another example. If you ask a group of students to study the rivers on the borderline between Iran and Iraq, and to write about them, they will present sentences similar to those we mentioned in the fifth section. You may afterwards ask them a question like "Can a river originate from another river?" The possible answers of the students can be put into three categories: 1. Affirmative; 2. Negative; and 3. Uncertain (e.g., "I don't know."). In all the three cases, students look for a sample in their memory. Some will find combinations such as Tigris, Karun, and Shatt al-Arab in real world and therefore respond in the affirmative. Some will not retrieve any such example in their memory about the real world and therefore will say "I don't know" in a very realistic manner. And some will respond in the negative because, on the one hand, they are not aware of such a possibility which is in its own turn due to their inability to recall any such instance in the real world, and, on the other hand, because they are confident about their knowledge, which differentiates them from the members of the previous group. In all three cases, human learning has been based on instances from the real world. Such questions in our proposed method are answered with help of collective knowledge which here is Linked Data. It is clear that questions such as "Can a river originate from another river?" are among those which semantic web can provide answer to. In Linked data RDF triples are collected so that such questions can be answered. Therefore our proposed approach would collect instances from text and put the answers to such questions in intermediate ontology. Obviously, the ontology's reasoning power becomes stronger. Such a process has never been put forth in any of the eight layers of ontology learning stack.

Another aspect of the proposed approach is as follows. Generally Linked Data is way to describe structured data [1, 2 and 14]. For instance structured data can be data existing in databases which have meanings of their own in the storage structure – tables, limitations on tables, tables' relations, etc. in a relational database. This storage structure actually reveals the designer's and analyst's understandings of the operational environment, entities and the relations between them these are another set of hidden information. In contrast to the approaches to ontology learning from pure text, ontology creation or enrichment based on Linked Data can take advantage of this hidden information. If the intermediate ontology is created from text and the Linked Data, in the same domain, is created from a database, this hidden information can definitely help enrich the intermediate ontology.

Also we can use inductive reasoning in our enrichment process to get a better result. The example that we prepared is an evidence of this claim.

Our proposed approach inherits dynamism from Linked Data; however the current LOD cloud is not a suitable base for our proposal in all informational domains. The reason we chose the geographical domain as an illustrating example is the abundance of the geographical resources in Linked Data. The more informational domains covered in the LOD cloud, the more obvious the importance of our proposed approach.

## VIII. CONCLUSION

In this paper we propose a novel approach for extracting some hidden assertional knowledge from text by using proper Linked Data dataset which results in achieving new Ontological Knowledge. We use Linked Data as collective knowledge to make use of hidden or implied information in texts, from which new ontological relations can be inferred. We showed that using Linked Data can improve the problem of context-awareness in the case of automatic ontology learning process. In this context, we proposed an algorithm to make use of Linked Data to enrich the non-taxonomic relations in the ontologies extracted from texts. We illustrated that this algorithm can find new non-taxonomic relations. We also show the soundness of our algorithm by using a real example in geographical domain. To trace our algorithm, we have searched for new predicates in FactForge.net We, also, have illustrated the possibility of this process by performing our algorithm on a real example which uses current Linked Data.

## IX. FUTURE WORK

As our future work we are planning to select and extend an algorithm to check the similarity of contexts and we will complete our system and evaluate it with other datasets. Furthermore, we want to present a definition for "enrichment extremity" based on the capacity and limitations of the intermediate ontology and limitations of Linked Data. Also we want to evaluate the claim that from the word co-occurrence aspect our method for relation extraction results in lower obsolescent of information in comparison to current existing relation extraction methods. At the end we want to propose an algorithm that uses inductive reasoning in an effective manner to help enriching process.

Our point of view to the obsolescent of information in Linked Data is as follows. Lack of discovery of relations between two instances, that is less enrichment, is because of obsolescent of relations in Linked Data. This also has two other reasons by itself. A) Little growth of Linked Data in comparison to the amount of existing data in traditional web. B) Even if the growth percentage of becomes more than it is, also there exists the problem of obsolescent of thoughts and ontologies in Linked Data. We think that this is because of the thought that the current Linked Data is the product of best practices. So we want to determine some metric to better describe the problem of obsolescent of information in Linked Data.